\documentclass{./styles/svproc}
\usepackage{booktabs}
\usepackage{amsmath}
\usepackage{amsfonts}
\usepackage{amssymb}
\usepackage{graphicx}
\usepackage{appendix}
\usepackage[normalem]{ulem}
\usepackage[misc]{ifsym}

\usepackage{url}

\begin{document}
\mainmatter              
%
\title{A Case Study of Agent-Based Models for Evolutionary Game Theory}
%
%
\author{Jacobus Smit\inst{1} \Letter \and Edward Plumb\inst{2}}
\authorrunning{J. Smit \and E. Plumb}
\institute{University of Oxford: \email{jacobus.smit@stx.ox.ac.uk},\\
\and
London School of Economics: \email{e.plumb@lse.ac.uk},\\}

\maketitle              

\begin{abstract}
Evolutionary game theory is a mathematical toolkit to analyse the interactions that an individual agent has in a population and how the composition of strategies in this population evolves over time. While it can provide neat solutions to simple problems, in more complicated situations where assumptions such as infinite population size may be relaxed, deriving analytic solutions can be intractable. In this short paper, we present a game with complex interactions and examine how an agent-based model may be used as a heuristic technique to find evolutionarily stable states.

\end{abstract}
\section{Introduction}
The treatment of evolutionary game theory (EGT) by agent-based modelling (ABM) is covered comprehensively in \cite{egtusingabm} which rightfully states that a mathematical treatment of EGT problems sometimes makes unreasonable assumptions about the problem such as infinite population size. It concludes that ``while [ABM] can tread where mathematics cannot'', a mathematical approach should be taken to validate the results by describing the limiting cases of ABM.

Our paper makes a similar argument with a different approach. Rather than focusing on the limitations of mathematical models in their ability to express more complex properties of problems (such as stochastic decisions and communication), we argue that, from a computational complexity perspective, the use of ABM can expedite the identification of potential evolutionarily stable states (ESS). This can then be validated using mathematics to arrive at the same conclusion as before.

This is done by the introduction, solution, and discussion of a specific EGT problem. We solve it both analytically and show that the ABM becomes close to the analytic model the more times it is run. We also discuss the choices we made in our ABM that are relevant to our argument that ABM provides a computationally efficient way of heuristically solving EGT problems. Further details of the implementation of the model can be found in the appendix.

\section{Problem Setup}
The example we will work with comes from Dawkins' \textit{The Selfish Gene} \cite{selfishgene}. It discusses how certain birds have evolved the strategy of sneaking their eggs into other birds' nests, so they don't have to expend the energy to look after the eggs themselves. Cuckoos are an example of birds (referred to as brood parasites or, in this paper, ``cheaters'') that perform this strategy.

However, certain species of birds (hereon ``identifiers'') have evolved to be able to identify their own eggs \cite{egg-recognition}. This means that a cheater unfortunate enough to lay their egg in one of these nests would not get any benefit from doing so. Guillemots are birds who can identify their own eggs; on the other hand, this identification process may take a while and use energy that could otherwise be used on finding food. Finally, some species such as chickens (hereon ``sitters'') have such a strong ingrained instinct to look after whatever is in their nest that they will sit on anything that falls into it, including fake eggs and even small animals such as kittens.

We assume that birds lay only a single egg (which can abstractly represent their offspring as a whole), sitting on an egg always causes it to hatch a chick, and that cheaters choose which nest to lay their eggs in uniformly at random amongst the set of nests. For every egg that hatches, the bird who laid the egg will get $h$ utility. For every egg that the bird has to sit on, the bird will lose $e$ utility. If a bird has to take the time to identify its own eggs out of all those in its nest it will lose $i$ utility. As in real life we assume that cheaters do not build nests, and so do not sit on eggs.

This problem provides a situation involving 3 distinct agent-types, an element of stochasticity, a multitude of tuneable parameters, and multiple directions of expansion in complexity. While it is solvable analytically, the elegance of the ABM in handling the above features, and its ease of extensibility will ultimately make ABM the more attractive method.

\begin{figure}[h]
\centering
\caption{Summary of the payoffs}
    \begin{tabular}{@{}ll@{}}
    \\\toprule
        Situation                 & Reward/Cost \\ \midrule
        Own egg hatches &   $h$           \\
        Sitting on eggs           & $-e$ per egg\\
        Identifying all eggs          & $-i$ \\\bottomrule
    \end{tabular}
    \label{table:utility}
\end{figure}

\section{Analytic Solution}
Let $p_S, p_I, p_C \in [0, 1]$ where $p_S + p_I + p_C = 1$ be the proportion of `Sitters’, ‘Identifiers’ and ‘Cheaters’ in the population. Due to the stochastic nature of the game the concept of a Nash Equilibrium (NE) does not make sense as, due to randomness, there can be no ``best'' strategy when at an interior point.  We must instead look for Bayesian NE states, in which all of the strategies have the same expected payoff. Each strategy's expected payoffs are given by

\[
\begin{aligned} \mathbb{E}(S) &= h - e\left(1 + \frac{p_{C}}{p_{S}+p_{I}}\right)\\ \mathbb{E}(I) &= h - e - i\\ \mathbb{E}(C) &= h \frac{p_{S}}{p_{S}+p_{I}} \end{aligned}
\]
where if $p_{S}+p_{I} = 0$, we say $\mathbb{E}(S) = -\infty$, and $\mathbb{E}(C) = 0$.

\begin{proposition}
Given $h,e,i >0$ and $h-e-i>0$, the game described in the previous section has a single (B)NE which is not an ESS at:
$$(p_S, p_I, p_C) = \left(\frac{e(h-e-i)}{h(i+e)}, \frac{e}{h}, \frac{i}{i+e}\right)$$
\end{proposition}

\section{Agent-Based Model}
Our aim was to make the simplest model possible, using a pre-established package (\verb|Agents.jl| \cite{agents.jl}) and making implementation decisions relevant to performance.

The model was constructed to mimic the EGT problem as closely as possible so they can be directly compared, as opposed to accurately representing real life. This is exemplified in our decision to use a utility-based proportional reproduction method as opposed to having eggs that hatch forming the new generation.

The model is initialised with parameters and agents, the agents act in an order corresponding to their type (as if identifiers acted before cheaters they would identify \textit{before} the eggs were laid in their nest), and then the model counts up the total utility of the agents and samples the new generation of each type from the performance of the old one.

\section{Results}



To see how closely our ABM models the EGT model, we can calculate the ABM's predictions for expected utility at any given point and compare it to the analytic expected payoff calculated earlier. We do so by repeatedly initialising a model at the point we want to analyse, and running it for a single step multiple times, each time collected the utility of each type. We can then average these collections and compute performance metrics. As the number of repetitions increases, the model will then converge to the true utilities.

\begin{figure}[h]
    \centering
    \includegraphics[width=0.85\textwidth]{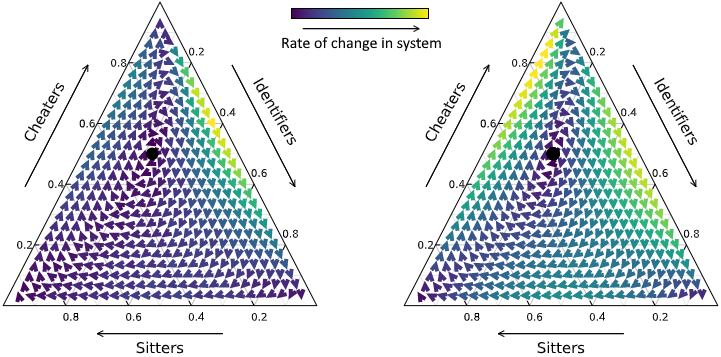}
    \caption{Left: Dynamics in the agent-based model each step repeated 150 times.\newline
        Right: Replicator dynamics for the EGT model. NE marked in black on both.}
    \label{fig:replicator_dynamics}
\end{figure}

We can also use our model to construct a vector field plot as in figure \ref{fig:replicator_dynamics} which shows the replicator dynamics of both the EGT model and the ABM.

\section{Discussion}
From figure \ref{fig:replicator_dynamics} we can see that our model does accurately predict the dynamics of the system and it is quite clear where the NE is on both graphs, even if that has not been calculated.

This particular model, however, is not the best demonstration of the power of ABM to model evolutionary games as it lacks an ESS. There are two computational complexity problems regarding ESS: their existence and their identification. It is known that the computational complexity of determining the existence of an ESS is \(\Sigma_2^p\text{-complete}\) \cite{ESScomplexity} and so heuristic methods could be very useful in situations where the problem cannot be simplified to a polynomial time solution. It is also known that given a candidate ESS, it is co\textbf{NP}-hard to check whether it is an ESS after all \cite{checkingcomplexity}. However, what is not known is the complexity of finding an ESS in the first place. An ABM could be used to quickly identify candidates that have basins of attraction or other promising features and methods of determining the location of the ESS could be concentrated in these areas, reducing the overall search space and making identification of ESS much faster. To use the ABM as a herustic method we can use a simple algorithm: run the model with some increasing number of agents and replicates until a potential ESS is identified then ``zoom in'' on this region and repeat until a desired accuracy is achieved. Ultimately, however, ABMs can perform strangely in corner cases which should be validated using mathematics.

In general agent-based models do require a lot of compute, but it is possible that models themselves can be parallelised. For example, in this example model the actions of the identifiers and sitters are predetermined and could be handled by the model, this means the agents do not need a specific order to act in and so agent actions could be distributed. In order to avoid resource contention it would be necessary to keep agents somewhat self contained, which can be difficult where more complex interactions are required. In this model the main interactions between agents come from the cheaters who place their agent ID in other agents' hash maps in order to store where their eggs are laid. This could result in contention if cheaters on multiple cores attempt to lay their eggs in the same nest. To solve this, the cheater could instead keep its own record of which nest to lay its egg into, and these records could be combined at the model level.

Finally we note that various types of hash maps are central to the implementation of \verb|Agents.jl| which, while a very useful and convenient tool, are not available for GPU computing. As such, even if a model could be parallelised for CPUs, a purely array based method would be required for implementation on GPUs.

%
%

\appendix
\section{Implementation Details}

Using \verb|Agents.jl| for multi-population systems there are two ways to approach implementing different types of agent. The most intuitive way is to assign a new mutable type for each agent, and populate the model with all of the different types. Another approach is to instead have only one agent type with the attributes and actions of every type, and an additional variable indicating its actual type. The conceptual advantage of doing the latter being that the concept of a mixed-strategy can be hard-coded into an agent. Computationally, the second approach is also (in current versions of \verb!Agents.jl!) faster due to type stability. We decided to go with the former method for our model for simplicity but recognise that the latter is superior.

Secondly, in EGT given each time step there must be a reproductive process such that the population can update itself, and be given the opportunity to mutate. The concept of utility in the original problem is interpreted as ``reproductive ability'', but the way an agent is awarded utility is precisely for hatching an egg i.e. reproducing. This means that utility is less of an ex-ante predictor of reproductive ability but rather an ex-post assessment of how an individual performed in the current time step. From this there are two ways of implementing reproduction. Naturally, it would make sense to equate an egg hatched with an agent in the following time step, but then the model strays further from the original EGT setup where there is no clear role for $h$. Additionally, one must consider the mortality of agents. If agents are hatched into existence then there also needs to be a way of removing them otherwise the number of agents in the model will explode. In the EGT setup, there is no problem of this sort as changes in population are calculated using the replicator dynamics \cite{egt-lecturenotes}. Our model uses utilities and proportional reproduction (where the current generation's agents are removed and new agents are sampled proportionally based on the performance of each type in the current generation) in the hope that this would make results from the analytic methods correspond better to those of the model, and made comparisons easier.

\end{document}